\documentclass[aps,preprint,showpacs,preprintnumbers,amsmath,amssymb]{revtex4}
\usepackage{amsmath,mathrsfs,amsbsy,color,graphicx,bm,amsthm,amsfonts}
\usepackage{units}
\usepackage{bbm}
\usepackage{times}
\usepackage{dcolumn}
\usepackage{mathrsfs}
\usepackage{amsmath,amssymb,epsfig}
\usepackage{amsmath}
\newcommand{\udots}{\mathinner{\mskip1mu\raise1pt\vbox{\kern7pt\hbox{.}}
\mskip2mu\raise4pt\hbox{.}\mskip2mu\raise7pt\hbox{.}\mskip1mu}}
\begin{document}
\title{ Do maximally entangled states always have an advantage over non-maximally entangled states in Schwarzschild black hole? }
\author{Shu-Min Wu$^1$\footnote{Email: smwu@lnnu.edu.cn}, Si-Han Li$^1$}
\affiliation{$^1$ Department of Physics, Liaoning Normal University, Dalian 116029, China}


\begin{abstract}
It is generally believed that quantum entanglement in the maximally entangled states is greater than quantum entanglement in the non-maximally entangled states under a relativistic setting.
In this paper, we study quantum entanglement for  four different types of Bell-like states of the fermionic modes near the event horizon of a Schwarzschild black hole.
It is interesting to find that quantum entanglement in the maximally entangled states  is less than quantum entanglement in the non-maximally entangled states in Schwarzschild spacetime.
From the perspective of quantum resources, the non-maximally entangled states may have more advantages in curved spacetime compared to the maximally entangled states. This is obviously different from the conclusions in previous paper. For two types of Bell-like states, quantum entanglement suffers sudden death under the Hawking effect of the black hole, and for the other two types of Bell-like states, quantum entanglement can exist forever regardless of the Hawking temperature.  Therefore, we should choose appropriate types of  Bell-like states to handle relativistic quantum information tasks.
\end{abstract}

\vspace*{0.5cm}
\maketitle
\section{Introduction}
From Einstein's perspective, the gravitational collapse of sufficiently massive stars would form intriguing black holes, which are fascinating entities in our universe \cite{Q1,Q2}.
According to the no-hair theorem, the black hole retains only information about its mass, charge, and angular momentum, but loses all other details \cite{Q3}. Recent advances in astronomy have provided important direct and indirect evidence for these mysterious phenomena, most notably the first detection of gravitational waves by the Advanced LIGO and Virgo detectors, a discovery that marked an important milestone and revealed the merging process of binary black hole systems \cite{Q4}. In addition, the Event Horizon Telescope (EHT) achieved a major achievement: the first successful image of the supermassive black hole at the center of the giant elliptical galaxy $M87$ \cite{Q5,Q6,Q7,Q8,Q9,Q10}. At the same time, the EHT also successfully imaged another supermassive black hole, located in $Sgr$ $A^*$ \cite{Q11}. Because black holes are so far away from us and have special properties, they have always been shrouded in mystery and have become one of the frontier research fields, such as the black hole information loss paradox \cite{Q13,Q14,ZQ14}. Hawking predicted that vacuum fluctuations near the event horizon would cause the black hole to evaporate. In other words, particle-antiparticle pairs would be generated near the event horizon. Hawking speculated that antiparticles would fall into the black hole, while particles would escape from the black hole. Obviously, the Hawking radiation is the root cause of the black hole information loss paradox. Quantum entanglement, particularly influenced by Hawking radiation, represents a potentially crucial means of addressing the black hole information loss paradox.

Quantum information in the gravitational background, which merges quantum information, quantum field theory, and gravity, primarily focuses on two areas of research: (i) utilizing quantum technology to investigate the structure of spacetime; (ii) examining how gravitational effects influence quantum resources. There has been extensive research into how gravitational effects  influence quantum steering, entanglement,  discord, coherence, and entropic uncertainty relations \cite{Q15,Q16,Q17,Q18,Q19,Q20,Q21,Q22,Q23,Q24,Q25,Q26,Q27,Q28,Q29,Q30,Q31,Q32,Q33,Q34,Q35,Q36,Q37,Q38,Q39,Q40,tQ40,tQ41,tQ42,tQ43,tQ44}.
These studies have shown that the quantum resources of the maximally entangled states are more advantageous than those of the non-maximally entangled states in a relativistic setting \cite{Q26,Q27,Q28,Q29,Q30,Q31,Q32,Q33,Q34,Q35,Q36,Q37,Q38,Q39,Q40,tQ40}. However, it is more challenging to prepare maximally entangled states compared to non-maximally entangled states in the experiment. Consequently, non-maximally entangled states are often used in relativistic quantum information tasks. This raises the question: Are quantum resources in the non-maximally entangled states more advantageous than those in the maximally entangled states under a relativistic setting? This question motivates our research. Additionally, we aim to explore whether gravitational effects influence quantum entanglement of four different types of Bell-like states in the same way.  Our goal is to identify the Bell-like state with the largest residual entanglement under gravitational influence, which serves as another motivation for our research.

In this study, we will study the impact of the Hawking effect on quantum entanglement for four different types of Bell-like states of fermionic fields in Schwarzschild spacetime. Initially, we assume that Alice and Bob share the Bell-like states of Dirac fields in flat Minkowski spacetime. Subsequently, Alice and Bob hover near the event horizon of the black hole. We will calculate fermionic entanglement and derive the analytical expressions for four different types of Bell-like states in curved spacetime. We will then investigate how the Hawking effect from the black hole affects the degradation of fermionic entanglement between Alice and Bob. Our study yields two intriguing conclusions: (i) quantum entanglement of the non-maximally entangled states can be greater than that of the maximally entangled states in the Schwarzschild black hole;  (ii) the Hawking effect can cause quantum entanglement to experience sudden death for two types of Bell states, while for the other two types of Bell states, quantum entanglement can persist indefinitely in curved spacetime.

The paper is organized as follows. In Sec.II, we discuss the quantization of Dirac fields in the background of a Schwarzschild black hole.  In Sec.III, we study fermionic entanglement for four different types of Bell-like states  in the background of the Schwarzschild black hole.  Finally the summary is arranged in Sec.IV.

\section{Quantization of Dirac fields in Schwarzschild spacetime }

The metric of the Schwarzschild black hole \cite{Q22}  is found to be
\begin{eqnarray}\label{w11}
ds^2&=&-(1-\frac{2M}{r}) dt^2+(1-\frac{2M}{r})^{-1} dr^2\nonumber\\&&+r^2(d\theta^2
+\sin^2\theta d\varphi^2),
\end{eqnarray}
where the parameters $r$ and $M$ are represented by the radius and mass of the Schwarzschild black hole,  respectively.
For the sake of simplicity, we regard $\hbar, G, c$ and $k$ as unity in this paper.
The Dirac equation $[\gamma^a e_a{}^\mu(\partial_\mu+\Gamma_\mu)]\Phi=0$ \cite{Q41}  in Schwarzschild spacetime can be given by the following equation
\begin{eqnarray}\label{w12}
&&-\frac{\gamma_0}{\sqrt{1-\frac{2M}{r}}}\frac{\partial \Phi}{\partial t}+\gamma_1\sqrt{1-\frac{2M}{r}}\bigg[\frac{\partial}{\partial r}+\frac{1}{r}+\frac{M}{2r(r-2M)} \bigg]\Phi \nonumber\\
&&+\frac{\gamma_2}{r}(\frac{\partial}{\partial \theta}+\frac{\cot \theta}{2})\Phi+\frac{\gamma_3}{r\sin\theta}\frac{\partial\Phi}{\partial\varphi}=0,
\end{eqnarray}
where $\gamma_i$ ($i=0,1,2,3$) are the  Dirac matrices \cite{Q27,Q35}.

By solving Eq.(\ref{w12}), we can obtain a set of positive (fermions) frequency outgoing solutions inside and outside the regions of the event horizon as
\begin{eqnarray}\label{w13}
\Phi^+_{{\bold k},{\rm in}}\sim \phi(r) e^{i\omega u},
\end{eqnarray}
\begin{eqnarray}\label{w14}
\Phi^+_{{\bold k},{\rm out}}\sim \phi(r) e^{-i\omega u},
\end{eqnarray}
where $\phi(r)$ represents four-component Dirac spinor and $u=t-r_{*}$ with the tortoise coordinate $r_{*}=r+2M\ln\frac{r-2M}{2M}$. Here, $\bold k$ and $\omega$ are the wave vectors and frequencies, which satisfy $|\mathbf{k}|=\omega$ in the massless Dirac field.
Based on Eqs.(\ref{w13}) and (\ref{w14}),  we can expand the Dirac field $\Phi$ as
\begin{eqnarray}\label{w15}
\Phi&=&\int
d\bold k[\hat{a}^{\rm in}_{\bold k}\Phi^{+}_{{\bold k},\text{in}}
+\hat{b}^{\rm in\dag}_{\bold k}
\Phi^{-}_{{\bold k},\text{in}}\nonumber\\ &+&\hat{a}^{\rm out}_{\bold k}\Phi^{+}_{{\bold k},\text{out}}
+\hat{b}^{\rm out\dag}_{\bold k}\Phi^{-}_{{\bold k},\text{out}}],
\end{eqnarray}
where $\hat{a}^{\rm in}_{\bold k}$ and $\hat{b}^{\rm in\dag}_{\bold k}$ are the fermion annihilation and antifermion creation operators  inside the event horizon, and $\hat{a}^{\rm out}_{\bold k}$ and $\hat{b}^{\rm out\dag}_{\bold k}$ are the fermion annihilation and antifermion creation operators in the external region of the event
horizon, respectively. These annihilation and creation operators satisfy the canonical anticommutation rule $\{\hat{a}^{\rm out}_{\mathbf{k}},\hat{a}^{\rm out\dagger}_{\mathbf{k'}}\}=
\{\hat{b}^{\rm in}_{\mathbf{k}},\hat{b}^{\rm in\dagger}_{\mathbf{k'}}\}
=\delta_{\mathbf{k}\mathbf{k'}}. $  We  can  use $\hat{a}^{\rm in}_{\bold k}|0\rangle_S=\hat{a}^{\rm out}_{\bold k}|0\rangle_S=0$ to define the Schwarzschild vacuum. Usually, the modes  $\Phi^\pm_{{\bold k},{\rm in}}$ and $\Phi^\pm_{{\bold k},{\rm out}}$ are known as Schwarzschild modes.

Based on the suggestions by Domour and Ruffini \cite{Q42}, Kruskal modes can be employed to make analytic continuations for Eqs.(\ref{w13}) and (\ref{w14}). However, the Kruskal observer is free to excite any accessible modes. Therefore, the single-frequency Kruskal mode cannot be directly mapped to a corresponding group of single-frequency Schwarzschild modes. To resolve this discrepancy, we introduce the Unruh mode, which serves as an intermediate bridge between the Kruskal and Schwarzschild modes \cite{Q43, Q44}. The Unruh operators exhibit simple Bogoliubov transformations with the Schwarzschild modes, which is given by
\begin{eqnarray}\label{w19}
\nonumber \tilde{c}_{\bold k,R}&=&\frac{1}{\sqrt{e^{-\frac{\omega}{T}}+1}}\hat{a}^{\rm out}_{\bold k}-\frac{1}{\sqrt{e^{\frac{\omega}{T}}+1}}\hat{b}^{\rm in\dag}_{\bold k},
\end{eqnarray}
\begin{eqnarray}\label{w20}
\nonumber \tilde{c}_{\bold k,L}&=&\frac{1}{\sqrt{e^{-\frac{\omega}{T}}+1}}\hat{a}^{\rm in}_{\bold k}-\frac{1}{\sqrt{e^{\frac{\omega}{T}}+1}}\hat{b}^{\rm out\dag}_{\bold k},
\end{eqnarray}
\begin{eqnarray}\label{ww19}
\nonumber \tilde{c}^{\dag}_{\bold k,R}&=&\frac{1}{\sqrt{e^{-\frac{\omega}{T}}+1}}\hat{a}^{\rm out\dag}_{\bold k}-\frac{1}{\sqrt{e^{\frac{\omega}{T}}+1}}\hat{b}^{\rm in}_{\bold k},
\end{eqnarray}
\begin{eqnarray}\label{ww20}
\tilde{c}^{\dag}_{\bold k,L}&=&\frac{1}{\sqrt{e^{-\frac{\omega}{T}}+1}}\hat{a}^{\rm in\dag}_{\bold k}-\frac{1}{\sqrt{e^{\frac{\omega}{T}}+1}}\hat{b}^{\rm out}_{\bold k},
\end{eqnarray}
where $T=\frac{1}{8\pi M}$ represents the Hawking temperature. Here, the subscripts $R$ and $L$ denote the ``right" and ``left" modes, respectively. Using
the operator ordering $\hat{a}^{\rm out}_{\bold k} \hat{b}^{\rm in}_{\bold k} \hat{b}^{\rm out}_{\bold k} \hat{a}^{\rm in}_{\bold k}$ , the vacuum for the Unruh mode is written as
\begin{eqnarray}\label{ww21}
\nonumber |0\rangle_U=&\frac{1}{{e^{-\frac{\omega}{T}}+1}}|0000\rangle
-\frac{1}{\sqrt{e^{\frac{\omega}{T}}+e^{-\frac{\omega}{T}}+2}}|0101\rangle\\
&+\frac{1}{\sqrt{e^{\frac{\omega}{T}}+e^{-\frac{\omega}{T}}+2}}|1010\rangle
-\frac{1}{{e^{\frac{\omega}{T}}+1}}|1111\rangle,
\end{eqnarray}
where $|nn'n''n'''\rangle \equiv |n\rangle_{\rm out}^{+} |n'\rangle_{\rm in}^{-} |n''\rangle_{\rm out}^{-} |n'''\rangle_{\rm in}^{+}$.
Here, $\{|n\rangle_{\rm out}^{+}\}$ and $\{|n\rangle_{\rm in}^{-}\}$ form orthonormal bases for the outside and inside regions of the Schwarzschild black hole, respectively \cite{Q43, Q44}. The superscript $\{+,-\}$ denote the fermion and antifermion, respectively.
The  excited state for the Unruh mode can be expanded as
\begin{eqnarray}\label{www21}
\nonumber |1\rangle_U=&[q_{R}(\tilde{c}^{\dag}_{\bold k,R} \bigotimes I_{L})+q_{L}(I_{R} \bigotimes \tilde{c}^{\dag}_{\bold k,L})]|0\rangle_{U}\\
\nonumber =& q_{R}|1_{\bold k}\rangle_R \bigotimes |0_{\bold k}\rangle_L +q_{L}|0_{\bold k}\rangle_R \bigotimes |1_{\bold k}\rangle_L \\
\nonumber =&q_{R}[\frac{1}{\sqrt{e^{-\frac{\omega}{T}}+1}}|1000\rangle
-\frac{1}{\sqrt{e^{\frac{\omega}{T}}+1}}|1101\rangle]\\
&+q_{L}[\frac{1}{\sqrt{e^{-\frac{\omega}{T}}+1}}|0001\rangle
+\frac{1}{\sqrt{e^{\frac{\omega}{T}}+1}}|1011\rangle],
\end{eqnarray}
with $|q_{R}|^{2}+|q_{L}|^{2}=1$. The Hawking radiation is generated by quantum fluctuations near the event horizon, where pairs of fermions and antifermions are spontaneously created. The fermion and antifermion can radiate toward either the inside or the outside regions, with the total probability $|q_{R}|^{2}+|q_{L}|^{2}=1$.

In the references \cite{Q16,Q22,Q30,Q36}, the initial vacuum state for the single-mode approximation  is considered to be  $$|0\rangle_K=\sum^1_{n=0}A_{n} |n\rangle_{\rm out} |n\rangle_{\rm in},$$
where $\{|n\rangle_{\rm out}\}$ and $\{|n\rangle_{\rm in}\}$ represent the Schwarzschild number states for the fermion outside the event horizon and the antifermion inside the horizon, respectively.  This corresponds to the case
$q_{R}=1$, where all fermions escape outward, while antifermions are confined within the horizon.
Therefore, after normalizing the state vector appropriately,  the Kruskal vacuum state for the single-mode approximation ($q_{R}=1$) can be represented in the Schwarzschild Fock space as
\begin{eqnarray}\label{w21}
 |0\rangle_K&=&\frac{1}{\sqrt{e^{-\frac{\omega}{T}}+1}}|0\rangle_{\rm out} |0\rangle_{\rm in}+\frac{1}{\sqrt{e^{\frac{\omega}{T}}+1}}|1\rangle_{\rm out} |1\rangle_{\rm in}.
\end{eqnarray}
Applying the operator $\hat{C}^{\rm out\dag}_{\bold k}=\frac{1}{\sqrt{e^{-8\pi M\omega}+1}}\hat{a}^{\rm out\dag}_{\bold k}-\frac{1}{\sqrt{e^{8\pi M\omega}+1}}\hat{b}^{\rm in}_{\bold k}$ to the vacuum state yields the excited state in Schwarzschild Fock space
\begin{eqnarray*}
\begin{split}
|1\rangle_K=&\hat{C}^{\rm out\dag}_{\bold k} |0\rangle_{K}=(\frac{1}{e^{-8\pi M\omega}+1}\hat{a}^{\rm out\dag}_{\bold k}-\frac{1}{e^{8\pi M\omega}+1}\hat{b}^{\rm in}_{\bold k}\hat{a}^{\rm out\dag}_{\bold k}\hat{b}^{\rm in\dag}_{\bold k})|0\rangle_{\rm out} |0\rangle_{\rm in}\\
=&(\frac{1}{e^{-8\pi M\omega}+1}\hat{a}^{\rm out\dag}_{\bold k}+\frac{1}{e^{8\pi M\omega}+1}\hat{a}^{\rm out\dag}_{\bold k}\hat{b}^{\rm in}_{\bold k}\hat{b}^{\rm in\dag}_{\bold k})|0\rangle_{\rm out} |0\rangle_{\rm in}\\
=&\hat{a}^{\rm out\dag}_{\bold k}|0\rangle_{\rm out} |0\rangle_{\rm in}=|1\rangle_{\rm out} |0\rangle_{\rm in}.
\end{split}
\end{eqnarray*}
Regarding the definition of the operator $\hat{C}^{\rm out\dag}_{\bold k}$, we consider that it is not equivalent to the standard annihilation operator defined via the complete Kruskal mode expansion. In the context of relativistic quantum information, particularly under the single-mode approximation, $\hat{C}^{\rm out\dag}_{\bold k}$ is often constructed as a linear combination of operators with support predominantly outside the horizon, as also done in previous papers \cite{Q15,Q16,Q22,Q30}. While this approach does not strictly follow the full mode expansion formalism, it has proven useful for capturing essential features of entanglement degradation and quantum correlations near black holes. Therefore, the one-particle state
$|1\rangle_K$ used in our work should not be mistaken for the standard one constructed via global mode support.

Near the event horizon of a black hole, quantum fluctuations in the vacuum produce particle-antiparticle pairs. This process does not distinguish between positive and negative particles, but is simply a spontaneous fluctuation of the quantum vacuum in the extremely strong background of gravity. What may happen next is that one particle of the pair falls into the black hole, and the other escapes to infinity, which is seen as Hawking radiation by distant observers. This escaping particle can be either a particle or an antiparticle. Both situations are possible with equal probability. In the single-mode approximation commonly adopted in relativistic quantum information, it is usually assumed that all antiparticles fall into the black hole while all particles escape to infinity. This convention is not a fundamental physical principle, but rather a modeling simplification. It enables tractable calculations of entanglement and information measures by effectively reducing the full multimode structure to a two-mode system \cite{Q15,Q16,Q22,66Q44}. The assignment of particles and antiparticles to exterior and interior regions is coordinate-dependent and serves primarily to facilitate the construction of Bogoliubov-transformed states under specific observer perspectives.

The Schwarzschild observer hovers outside the event horizon and its Hawking radiation spectrum is given by \cite{Q16,Q36}
\begin{eqnarray}\label{w22}
N_F=\sideset{_K}{}{\mathop{\langle}}0|\hat{a}^{\rm out\dag}_{\bold k}\hat{a}^{\rm out}_{\bold k}|0\rangle_K=\frac{1}{e^{\frac{\omega}{T}}+1}.
\end{eqnarray}
From the viewpoint of the Schwarzschild observer, this equation suggests that the Kruskal vacuum, as observed by them, would manifest as a count of generated fermions $N_F$. In simpler terms, the Schwarzschild observer outside the black hole can perceive a thermal distribution following Fermi-Dirac statistics.

\section{Quantum entanglement for  different types of Bell-like states in Schwarzschild spacetime}
Initially, we consider four different types of Bell-like states of the entangled fermionic modes in the asymptotically flat region of a Schwarzschild black hole
\begin{eqnarray}\label{w23}
|\phi^{{1},{\pm}}_{AB}\rangle=\alpha|0_{A}\rangle|0_{B}\rangle \pm \sqrt{1-\alpha^{2}}|1_{A}\rangle|1_{B}\rangle,
\end{eqnarray}
\begin{eqnarray}\label{w24}
|\Psi^{{2},{\pm}}_{AB}\rangle=\alpha|0_{A}\rangle|1_{B}\rangle \pm \sqrt{1-\alpha^{2}}|1_{A}\rangle|0_{B}\rangle,
\end{eqnarray}
where the subscripts $A$ and $B$ represent the modes  associated with Alice and Bob, respectively.
Subsequently, both Alice and Bob are positioned outside the event horizon of the black hole. Consequently, Alice and Bob will detect the thermal Fermi-Dirac statistics, and their detectors are found to be excited. Using Eq.(\ref{w21}),  we can rewrite Eqs.(\ref{w23}) and (\ref{w24})  in terms of Schwarzschild modes for Alice and Bob
\begin{eqnarray}\label{w25}
|\phi^{{1},{\pm}}_{A{\bar A}B{\bar B}}\rangle&=&
\alpha(
\cos r_{A}\cos r_{B}|0\rangle_{A}|0\rangle_{\bar A}|0\rangle_{B}|0\rangle_{\bar B}
+\cos r_{A}\sin r_{B}|0\rangle_{A}|0\rangle_{\bar A}|1\rangle_{B}|1\rangle_{\bar B}\nonumber\\&&
+\sin r_{A}\cos r_{B}|1\rangle_{A}|1\rangle_{\bar A}|0\rangle_{B}|0\rangle_{\bar B}
+\sin r_{A}\sin r_{B}|1\rangle_{A}|1\rangle_{\bar A}|1\rangle_{B}|1\rangle_{\bar B})\nonumber\\&&\pm
\sqrt{1-\alpha^{2}}|1\rangle_{A}|0\rangle_{\bar A}|1\rangle_{B}|0\rangle_{\bar B},
\end{eqnarray}
\begin{eqnarray}\label{w26}
|\Psi^{{2},{\pm}}_{A{\bar A}B{\bar B}}\rangle&=&
\alpha \cos r_{A}|0\rangle_{A}|0\rangle_{\bar A}|1\rangle_{B}|0\rangle_{\bar B}
+\alpha \sin r_{A}|1\rangle_{A}|1\rangle_{\bar A}|1\rangle_{B}|0\rangle_{\bar B}\nonumber\\&&\pm
\sqrt{1-\alpha^{2}}\cos r_{B}|1\rangle_{A}|0\rangle_{\bar A}|0\rangle_{B}|0\rangle_{\bar B}\pm
\sqrt{1-\alpha^{2}}\sin r_{B}|1\rangle_{A}|0\rangle_{\bar A}|1\rangle_{B}|1\rangle_{\bar B},
\end{eqnarray}
where the modes $\bar A$ and $\bar B$ are observed by hypothetical observers Anti-Alice and Anti-Bob inside the event horizon of the black hole.
For simplicity, we define $\cos r_{A}=\frac{1}{\sqrt{e^{-\frac{\omega_{A}}{T}}+1}}$,
$\sin r _{A}=\frac{1}{\sqrt{e^{\frac{\omega_{A}}{T}}+1}}$,
$\cos r_{B}=\frac{1}{\sqrt{e^{-\frac{\omega_{B}}{T}}+1}}$,
and $\sin r_{B}=\frac{1}{\sqrt{e^{\frac{\omega_{B}}{T}}+1}}$,
where $\omega_{A}$ and $\omega_{B}$ are the frequencies of modes $A$ and $B$, respectively.

Because the exterior region of the black hole is causally disconnected from its interior, Alice and Bob cannot detect physically inaccessible modes $\bar A$ and $\bar B$.  Consequently, we need to trace over these inaccessible modes to derive the density matrix for Alice and Bob
\begin{eqnarray}\label{w27}
 \rho^{{1},{\pm}}_{AB}=\left(\!\!\begin{array}{cccccccc}
\alpha^{2} \cos^{2}r_{A} \cos^{2}r_{B}&0&0&\pm\alpha\sqrt{1-\alpha^{2}} \cos r_{A} \cos r_{B}\\
0&\alpha^{2} \cos^{2}r_{A} \sin^{2}r_{B}&0&0\\
0&0&\alpha^{2} \sin^{2}r_{A} \cos^{2}r_{B}&0\\
\pm\alpha\sqrt{1-\alpha^{2}}\cos r_{A} \cos r_{B}&0&0&\alpha^{2} \sin^{2}r_{A} \sin^{2}r_{B}+1-\alpha^{2}
\end{array}\!\!\right),
\end{eqnarray}
\begin{eqnarray}\label{w28}
 \rho^{{2},{\pm}}_{AB}=\left(\!\!\begin{array}{cccccccc}
0&0&0&0\\
0&\alpha^{2} \cos^{2}r_{A}&\pm\alpha\sqrt{1-\alpha^{2}}\cos r_{A} \cos r_{B}&0\\
0&\pm\alpha\sqrt{1-\alpha^{2}}\cos r_{A} \cos r_{B}&(1-\alpha^{2}) \cos^{2}r_{B}&0\\
0&0&0&\alpha^{2} \sin^{2}r_{A}+(1-\alpha^{2})\sin^{2}r_{B}
\end{array}\!\!\right).
\end{eqnarray}

In this paper, we utilize the concurrence as a metric to quantify quantum entanglement in the Schwarzschild black hole. For a mixed two-qubit state, its concurrence can be defined as
\begin{eqnarray}\label{w29}
 C(\rho)=\max\{0,\lambda_{1}-\lambda_{2}-\lambda_{3}-\lambda_{4}\}, \quad \lambda_{i} \geq \lambda_{i+1} \geq 0 ,
\end{eqnarray}
where $\lambda_{i}$ represents the square roots of the eigenvalues of the matrix $\rho\widetilde{\rho}$ with the ``spin-flip" matrix $\widetilde{\rho} = (\sigma_{y}\otimes\sigma_{y})\rho^{\ast}(\sigma_{y}\otimes\sigma_{y})$.
Employing Eq.(\ref{w29}), we obtain the analytical expressions of the concurrence for Eqs.(\ref{w27}) and (\ref{w28}) as
\begin{eqnarray}\label{w30}
C(\rho^{{1},{\pm}}_{AB})=\max\bigg\{0,2\alpha^{2}\cos r_{A} \cos r_{B}(\frac{\sqrt{1-\alpha^{2}}}{\alpha}-\sin r_{A} \sin r_{B})\bigg\},
\end{eqnarray}
\begin{eqnarray}\label{w31}
C(\rho^{{2},{\pm}}_{AB})=\max\{0,2\alpha\sqrt{1-\alpha^{2}}\cos r_{A} \cos r_{B}\}.
\end{eqnarray}

\begin{figure}
\begin{minipage}[t]{0.5\linewidth}
\centering
\includegraphics[width=3.0in,height=5.2cm]{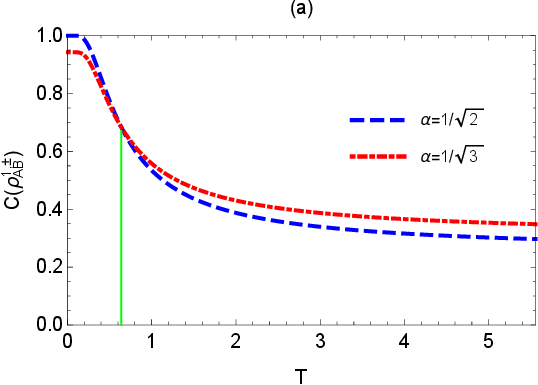}
\label{fig1a}
\end{minipage}%
\begin{minipage}[t]{0.5\linewidth}
\centering
\includegraphics[width=3.0in,height=5.2cm]{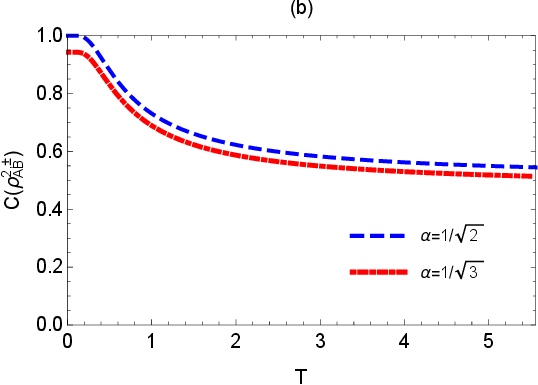}
\label{fig1c}
\end{minipage}%
\caption{The concurrence $C(\rho^{{1},{\pm}}_{AB})$ and $C(\rho^{{2},{\pm}}_{AB})$ as functions of the Hawking temperature $T$ for fixed $\omega_{A} = \omega_{B} = 1$.}
\label{Fig1}
\end{figure}

In Fig.\ref{Fig1}, we illustrate the variation of the concurrence $C(\rho^{{1},{\pm}}_{AB})$ and $C(\rho^{{2},{\pm}}_{AB})$ with the Hawking temperature $T$ for different types of Bell-like states. It is shown that, for $\alpha=\frac{1}{\sqrt{2}}$ and $\alpha=\frac{1}{\sqrt{3}}$, we find that the concurrence  first decreases and then approaches an asymptotic value as the Hawking temperature $T$ increases. From Fig.\ref{Fig1}(a),  we can see that for the initial Bell-like states $|\phi^{{1},{\pm}}_{AB}\rangle$,
the concurrence $C(\rho^{{1},{\pm}}_{AB})$ of the maximally entangled states $(\alpha=\frac{1}{\sqrt{2}})$ is greater than  the concurrence $C(\rho^{{1},{\pm}}_{AB})$ of the non-maximally entangled states ($\alpha=\frac{1}{\sqrt{3}}$) when $T < \frac{1}{\ln(2\sqrt{2}+2)}$ in Schwarzschild spacetime. However, the concurrence  of the maximally entangled state is smaller than  the concurrence of the non-maximally entangled state for $T > \frac{1}{\ln(2\sqrt{2}+2)}$. In other words, the non-maximally entangled states may have advantages over the maximally entangled states in curved spacetime. This is different from the previous conclusion that the maximally entangled states always have an advantage over the non-maximally entangled states in a relativistic setting \cite{Q26,Q27,Q28,Q29,Q30,Q31,Q32,Q33,Q34,Q35,Q36,Q37,Q38,Q39,Q40,tQ40}. Therefore, we should choose the non-maximally entangled states to handle relativistic quantum information tasks  in a strong gravitational environment.

From Eq.(\ref{w31}),  we find that for the initial Bell-like states $|\Psi^{{2},{\pm}}_{AB}\rangle$, the concurrence of the maximally entangled states  is always greater than  the concurrence  of the non-maximally entangled states in the Schwarzschild black hole. Fig.\ref{Fig1}(b) also bears witness to this conclusion. In addition, the concurrence for the initial Bell-like states $|\Psi^{{2},{\pm}}_{AB}\rangle$ is always greater than
the concurrence for the initial Bell-like states $|\phi^{{1},{\pm}}_{AB}\rangle$ under the Hawking effect of the black hole. This means that the Bell-like states $|\Psi^{{2},{\pm}}_{AB}\rangle$ are a better quantum resource than the Bell-like states $|\phi^{{1},{\pm}}_{AB}\rangle$ in a relativistic setting.

\begin{figure}
\begin{minipage}[t]{0.5\linewidth}
\centering
\includegraphics[width=3.0in,height=5.2cm]{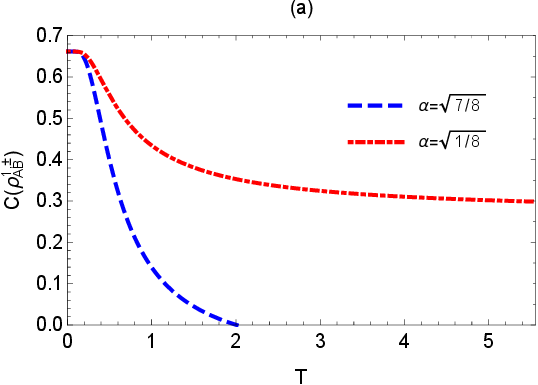}
\label{fig2a}
\end{minipage}%
\begin{minipage}[t]{0.5\linewidth}
\centering
\includegraphics[width=3.0in,height=5.2cm]{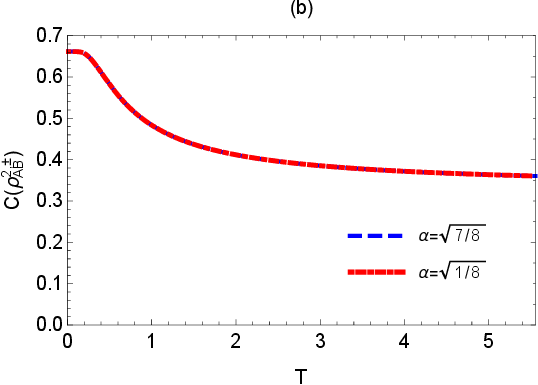}
\label{fig2b}
\end{minipage}%
\caption{The concurrence $C(\rho^{{1},{\pm}}_{AB})$ and $C(\rho^{{2},{\pm}}_{AB})$ as functions of the Hawking temperature $T$ for fixed $\omega_{A} = \omega_{B} = 1$.}
\label{Fig2}
\end{figure}

From Eq.(\ref{w30}), we can observe that for $\frac{\sqrt{1-\alpha^2}}{\alpha}>\sin r_{A} \sin r_{B}$, the concurrence $C(\rho^{{1},{\pm}}_{AB})$ can persist in Schwarzschild spacetime, otherwise it suffers sudden death. From Eq.(\ref{w31}), we can see that the concurrence $C(\rho^{{2},{\pm}}_{AB})$ can persist indefinitely for any Hawking temperature $T$. For $\alpha=\frac{1}{\sqrt{8}}$ and $\alpha=\sqrt{\frac{7}{8}}$, the initial states $|\phi^{{1},{\pm}}_{AB}\rangle$ have the same concurrence value in the asymptotically flat region.  However, the concurrence $C(\rho^{{1},{\pm}}_{AB})$ for $\alpha=\frac{1}{\sqrt{8}}$  is greater than the concurrence $C(\rho^{{1},{\pm}}_{AB})$ for $\alpha=\sqrt{\frac{7}{8}}$ under the influence of the Hawking effect. From Fig.\ref{Fig2}, we find that, with the increase of the Hawking temperature $T$,  the concurrence $C(\rho^{{1},{\pm}}_{AB})$ for $\alpha=\sqrt{\frac{7}{8}}$ experiences sudden death, while the concurrence $C(\rho^{{1},{\pm}}_{AB})$ for $\alpha=\frac{1}{\sqrt{8}}$ can survive forever. However, the initial states $|\Psi^{{2},{\pm}}_{AB}\rangle$ for $\alpha=\frac{1}{\sqrt{8}}$ and $\alpha=\sqrt{\frac{7}{8}}$ have the same concurrence value in curved spacetime. Therefore, it is imperative to select an appropriate initial entangled state to effectively address relativistic quantum information tasks.

It is noteworthy that Eq.(\ref{w23}) can be expressed in the form of a density matrix as follows
\begin{eqnarray}\label{ww23}
\begin{split}
\rho^{{1}{'},{\pm}}_{AB}=&\alpha^{2} |0_{A}0_{B}\rangle\langle0_{A}0_{B}| \pm \alpha\sqrt{1-\alpha^{2}}|0_{A}0_{B}\rangle\langle1_{A}1_{B}| \\
&\pm \alpha\sqrt{1-\alpha^{2}}|1_{A}1_{B}\rangle\langle0_{A}0_{B}|+(1-\alpha^{2})|1_{A}1_{B}\rangle\langle1_{A}1_{B}|.
\end{split}
\end{eqnarray}
From Eq.(\ref{ww23}), it is evident that the initial state is symmetric when $\alpha^{2} \leftrightarrow 1-\alpha^{2}$.
After rewriting Eq.(\ref{ww23}) based on Eq.(\ref{w21}) and tracing out the operator inside the event horizon of the black hole, the initial basis undergo the following transformations
\begin{eqnarray*}
\begin{split}
\alpha^{2}|0_{A}0_{B}\rangle\langle0_{A}0_{B}|\rightarrow&
\alpha^{2}(\cos^{2}r_{A}\cos^{2}r_{B}|0_{A}0_{B}\rangle_{\rm out}\langle0_{A}0_{B}|+
\cos^{2}r_{A}\sin^{2}r_{B}|0_{A}1_{B}\rangle_{\rm out}\langle0_{A}1_{B}|\\
&+\sin^{2}r_{A}\cos^{2}r_{B}|1_{A}0_{B}\rangle_{\rm out}\langle1_{A}0_{B}|+
\sin^{2}r_{A}\sin^{2}r_{B}|1_{A}1_{B}\rangle_{\rm out}\langle1_{A}1_{B}|),\\
(1-\alpha^{2})|1_{A}1_{B}\rangle\langle1_{A}1_{B}|\rightarrow&(1-\alpha^{2})|1_{A}1_{B}\rangle_{\rm out}\langle1_{A}1_{B}|,\\
\alpha\sqrt{1-\alpha^{2}}|0_{A}0_{B}\rangle\langle1_{A}1_{B}| \rightarrow& \alpha\sqrt{1-\alpha^{2}} \cos r_{A} \cos r_{B}|0_{A}0_{B}\rangle_{\rm out}\langle1_{A}1_{B}|,\\
\alpha\sqrt{1-\alpha^{2}}|1_{A}1_{B}\rangle\langle0_{A}0_{B}| \rightarrow& \alpha\sqrt{1-\alpha^{2}}\cos r_{A} \cos r_{B}|1_{A}1_{B}\rangle_{\rm out}\langle0_{A}0_{B}|.
\end{split}
\end{eqnarray*}
It is obvious that both initial basis $\alpha^{2}|0_{A}0_{B}\rangle\langle0_{A}0_{B}|$ and $(1-\alpha^{2})|1_{A}1_{B}\rangle\langle1_{A}1_{B}|$ become asymmetric in Schwarzschild spacetime, which explains why the concurrence $C(\rho^{{1},{\pm}}_{AB})$ changes under the exchange $\alpha^{2} \leftrightarrow 1-\alpha^{2}$.
Moverover, Eq.(\ref{w24}) can also be expressed as a density operator
\begin{eqnarray}\label{ww24}
\begin{split}
\rho^{{2}{'},{\pm}}_{AB}=&\alpha^{2} |0_{A}1_{B}\rangle\langle0_{A}1_{B}| \pm \alpha\sqrt{1-\alpha^{2}}|0_{A}1_{B}\rangle\langle1_{A}0_{B}| \\
&\pm \alpha\sqrt{1-\alpha^{2}}|1_{A}0_{B}\rangle\langle0_{A}1_{B}|+(1-\alpha^{2})|1_{A}0_{B}\rangle\langle1_{A}0_{B}|.
\end{split}
\end{eqnarray}
Similarly, from Eq.(\ref{ww24}), we observe that the initial state is symmetric when $\alpha^{2} \leftrightarrow 1-\alpha^{2}$.
 The initial basis of Eq.(\ref{ww24}) in Schwarzschild spacetime reveals the following transformations
\begin{eqnarray*}
\begin{split}
\alpha^{2}|0_{A}1_{B}\rangle\langle0_{A}1_{B}|\rightarrow&
\alpha^{2}(\cos^{2}r_{A}|0_{A}1_{B}\rangle_{\rm out}\langle0_{A}1_{B}|
+\sin^{2}r_{A}|1_{A}1_{B}\rangle_{\rm out}\langle1_{A}1_{B}|),\\
(1-\alpha^{2})|1_{A}0_{B}\rangle\langle1_{A}0_{B}|\rightarrow&
(1-\alpha^{2})(\cos^{2}r_{B}|1_{A}0_{B}\rangle_{\rm out}\langle1_{A}0_{B}|
+\sin^{2}r_{B}|1_{A}1_{B}\rangle_{\rm out}\langle1_{A}1_{B}|),\\
\alpha\sqrt{1-\alpha^{2}}|0_{A}1_{B}\rangle\langle1_{A}0_{B}| \rightarrow& \alpha\sqrt{1-\alpha^{2}}\cos r_{A} \cos r_{B}|0_{A}1_{B}\rangle_{\rm out}\langle1_{A}0_{B}|,\\
\alpha\sqrt{1-\alpha^{2}}|1_{A}0_{B}\rangle\langle0_{A}1_{B}| \rightarrow& \alpha\sqrt{1-\alpha^{2}}\cos r_{A} \cos r_{B}|1_{A}0_{B}\rangle_{\rm out}\langle0_{A}1_{B}|.
\end{split}
\end{eqnarray*}
Both initial basis  $\alpha^{2} |0_{A}1_{B}\rangle\langle0_{A}1_{B}|$ and $(1-\alpha^{2})|1_{A}0_{B}\rangle\langle1_{A}0_{B}|$ evolve from a single term to two terms, and their transformations maintain symmetry. Therefore, $\rho^{{2}{'},{\pm}}_{AB}$ for $\omega_{A}=\omega_{B}$ remains symmetrical under evolution when $\alpha^{2} \leftrightarrow 1-\alpha^{2}$.
In conclusion, our analysis demonstrates that the asymmetry in the evolution of Eq.(\ref{ww23}) and the symmetry in the evolution of Eq.(\ref{ww24}) lead to the observed differences in concurrence when $\alpha^{2} \leftrightarrow 1-\alpha^{2}$. This finding is consistent with the pattern illustrated in Fig.\ref{Fig2}, providing a clear explanation for the behavior of the system.

\section{Conclusions}
The effect of the Hawking effect on quantum entanglement of Dirac fields for four different types of Bell-like states in Schwarzschild spacetime is investigated. Our model consists of two subsystems, $A$ and $B$, which are observed by Alice and Bob, respectively, positioned outside the event horizon of the black hole. We find that,  for the initial Bell-like states $|\phi^{{1},{\pm}}_{AB}\rangle$, the concurrence of the maximally entangled states can be smaller than the concurrence of the non-maximally entangled states, indicating  that the non-maximally entangled states may offer advantages over the maximally entangled states in curved spacetime. This is different from the results in previous papers that  quantum resources of the maximally entangled states are superior to those of the non-maximally entangled states in a relativistic setting \cite{Q26,Q27,Q28,Q29,Q30,Q31,Q32,Q33,Q34,Q35,Q36,Q37,Q38,Q39,Q40,tQ40}. It is shown that,  for $\alpha=\frac{1}{\sqrt{8}}$ and $\alpha=\sqrt{\frac{7}{8}}$, the initial states $|\phi^{{1},{\pm}}_{AB}\rangle$ have the same concurrence value in the asymptotically flat region and  different concurrence values in Schwarzschild spacetime. We also find that, for the initial Bell-like states $|\Psi^{{2},{\pm}}_{AB}\rangle$, the  concurrence can survive for any Hawking temperature, while  for the initial Bell-like states $|\phi^{{1},{\pm}}_{AB}\rangle$, the Hawking effect can destroy the  concurrence under the condition  $\frac{\sqrt{1-\alpha^2}}{\alpha}\leq\sin r_{A} \sin r_{B}$. The above results should be significant for us to process relativistic quantum information tasks in the future.

\begin{acknowledgments}
This work is supported by the National Natural
Science Foundation of China (Grant Nos. 12205133), LJKQZ20222315, JYTMS20231051, and  the Special Fund for Basic Scientific Research of Provincial Universities in Liaoning under grant NO. LS2024Q002.	
\end{acknowledgments}



\begin{thebibliography}{99}
\bibitem{Q1}
K. Schwarzschild,  \"{U}ber das gravitationsfeld eines massenpunktes nach der einsteinschen theorie,  Sitzungsberichte der k\"{o}niglich preussischen Akademie der Wissenschaften,  189 (1916).

\bibitem{Q2}
S. W. Hawking, Breakdown of predictability in gravitational collapse, Phys. Rev. D {\bf14}, 2460  (1976).

\bibitem{Q3}
N. G\"{u}rlebeck, No-Hair theorem for black holes in astrophysical environments, Phys. Rev. Lett. {\bf114}, 151102  (2015).

\bibitem{Q4}
B. P. Abbott et al., Observation of gravitational waves from a binary black hole merger,  Phys. Rev. Lett. {\bf116}, 061102 (2016).

\bibitem{Q5}
The Event Horizon Telescope Collaboration, First $M87$ event horizon telescope results. I. the shadow of the supermassive black hole, Astrophys. J. Lett. {\bf875}, L1 (2019).

\bibitem{Q6}
 The Event Horizon Telescope Collaboration, First ${M}87$ event horizon telescope results. II.
array and instrumentation, Astrophys. J. Lett. {\bf875}, L2 (2019).

\bibitem{Q7}
The Event Horizon Telescope Collaboration, First ${M}87$ event horizon telescope results.
III. data processing and calibration, Astrophys. J. Lett. {\bf875}, L3 (2019).

\bibitem{Q8}
The Event Horizon Telescope Collaboration, First ${M}87$ event horizon telescope results. IV.
imaging the central supermassive black hole, Astrophys. J. Lett. {\bf875}, L4 (2019).

\bibitem{Q9}
The Event Horizon Telescope Collaboration, First ${M}87$ event horizon telescope results. V.
physical origin of the asymmetric ring, Astrophys. J. Lett. {\bf875}, L5 (2019).

\bibitem{Q10}
The Event Horizon Telescope Collaboration, First ${M}87$ event horizon telescope results. VI.
the shadow and mass of the central black hole, Astrophys. J. Lett. {\bf875}, L6 (2019).

\bibitem{Q11}
The Event Horizon Telescope Collaboration, First sagittarius ${A}^*$ event horizon telescope results. i. the shadow of the supermassive black hole in the center of the milky way,
Astrophys. J. Lett. {\bf930}, L12 (2022).



\bibitem{Q13}
H. Terashima, Entanglement entropy of the black hole horizon, Phys. Rev. D {\bf61}, 104016 (2000).

\bibitem{Q14}
L. Bombelli, R. K. Koul, J. Lee, and R. D. Sorkin, Quantum source of entropy for black holes, Phys. Rev. D
{\bf34}, 373 (1986).

\bibitem{ZQ14}
M. Cadoni, M. Oi, and A.P. Sanna, Evaporation and information puzzle for 2D
nonsingular asymptotically flat black holes, J. High Energy Phys. {\bf06}, 211 (2023).


\bibitem{Q15}
I. Fuentes-Schuller, and R. B. Mann, Alice falls into a black hole: Entanglement in non-inertial frames, Phys. Rev. Lett. {\bf95}, 120404 (2005).

\bibitem{Q16}
P. M. Alsing, I. Fuentes-Schuller, R. B. Mann, and T. E. Tessier, Entanglement of Dirac fields in non-inertial frames,Phys. Rev. A {\bf74}, 032326 (2006).

\bibitem{Q17}
S. M. Wu, R. D. Wang, X. L. Huang, Z. Wang, Does gravitational wave assist vacuum steering and Bell nonlocality?, J. High Energy Phys. {\bf07}, 155  (2024).

\bibitem{Q18}
D. Ahn, Unruh effect as a noisy quantum channel, Phys. Rev. A {\bf98}, 022308 (2018).



\bibitem{Q19}
M. Montero, E. Mart\'{\i}n-Mart\'{\i}nez, The entangling side of the Unruh-Hawking effect, J. High Energy Phys. {\bf07},  006 (2011).


\bibitem{Q20}
Z. Tian, J. Jing, Geometric phase of two-level atoms and thermal nature of de Sitter spacetime,
J. High Energy Phys. {\bf04}, 109  (2013).

\bibitem{Q21}
G. Adesso, I. Fuentes-Schuller, and M. Ericsson, Continuous variable entanglement sharing in non-inertial frames, Phys. Rev. A {\bf76}, 062112 (2007).

\bibitem{Q22}
E. Mart\'{\i}n-Mart\'{\i}nez, L. J. Garay, and J. Le\'{o}n, Unveiling quantum entanglement degradation near a Schwarzschild black hole, Phys. Rev. D
{\bf82}, 064006 (2010).

\bibitem{Q23}
B. N. Esfahani, M. Shamirzaie, and M. Soltani, Reduction of entanglement degradation in Einstein-Gauss-Bonnet gravity, Phys. Rev. D {\bf84}, 025024 (2011).

\bibitem{Q24}
W. C. Qiang, G. H. Sun, Q. Dong, and S. H. Dong, Genuine multipartite concurrence for entanglement of Dirac fields in noninertial frames, Phys. Rev. A {\bf98}, 022320 (2018).

\bibitem{Q25}
Q. Pan, J. Jing, Hawking radiation, entanglement and teleportation in
background of an asymptotically flat static black hole, Phys. Rev. D {\bf78}, 065015 (2008).

\bibitem{Q26}
J. Wang, Q. Pan, S. Chen, J. Jing, Entanglement of coupled massive scalar field in background of dilaton black hole, Phys. Lett. B {\bf677}, 186 (2009).

\bibitem{Q27}
J. Wang, Q. Pan, J. Jing, Entanglement redistribution in the Schwarzschild spacetime, Phys. Lett. B {\bf692},  202 (2010).

\bibitem{Q28}
S. Kanno, J. P. Shock,  J. Soda, Quantum discord in de Sitter space, Phys. Rev. D {\bf94}, 125014 (2016).

\bibitem{Q29}
S. M. Wu, C. X. Wang, D. D. Liu, X. L. Huang, H. S. Zeng, Would quantum coherence be increased by curvature effect in de Sitter space?, J. High Energy Phys. {\bf02}, 115  (2023).

\bibitem{Q30}
S. Banerjee, A. K. Alok, S. Omkar, R. Srikanth, Characterization of Unruh channel in the context of
open quantum systems, J. High Energy Phys. {\bf02}, 082  (2017).

\bibitem{Q31}
S. M. Wu, X. W. Fan, R. D. Wang, H. Y. Wu, X. L. Huang, H. S. Zeng, Does Hawking effect always degrade fidelity of quantum teleportation in Schwarzschild spacetime?, J. High Energy Phys. {\bf11}, 232 (2023).

\bibitem{Q32}
J. He, S. Xu, Y. Yu, L. Ye, Property of various correlation measures of open Dirac system
with Hawking effect in Schwarzschild space-time, Phys. Lett. B {\bf740},   322 (2015).

\bibitem{Q33}
J. Le\'{o}n, E. Mart\'{\i}n-Mart\'{\i}nez, Spin and occupation number entanglement of Dirac fields for noninertial observers, Phys. Rev. A {\bf80}, 012314  (2009).

\bibitem{Q34}
J. Chang, Y. Kwon, Entanglement behavior of quantum states of fermionic systems in an accelerated frame, Phys. Rev. A {\bf85}, 032302 (2012).

\bibitem{Q35}
S. M. Wu, X. W. Teng, J. X. Li, S. H. Li, T. H. Liu, J. C. Wang, Genuinely accessible and inaccessible entanglement in Schwarzschild black hole,  Phys. Lett. B {\bf848}, 138334  (2024).

\bibitem{Q36}
J. Kumar Basak, D. Giataganas, S. Mondal, W. Y. Wen, Reflected entropy and Markov gap in noninertial frames, Phys. Rev. D {\bf108}, 125009 (2023).

\bibitem{Q37}
T. Zhang,  X. Wang, S. M. Fei,  Hawking effect can generate physically inaccessible genuine tripartite nonlocality, Eur. Phys. J. C  {\bf83}, 607 (2023).

\bibitem{Q38}
T. Y. Wang, D. Wang, Entropic uncertainty relations in Schwarzschild space-time, Phys. Lett. B {\bf855},  138876 (2024).

\bibitem{Q39}
S. M. Wu, H. S. Zeng, Genuine tripartite nonlocality and entanglement in curved spacetime,  Eur. Phys. J. C {\bf82}, 4 (2022).

\bibitem{Q40}
S. Sen, A. Mukherjee,  S. Gangopadhyay, Entanglement degradation as a tool to detect signatures of modified gravity, Phys. Rev. D {\bf109}, 046012 (2024).

\bibitem{tQ40}
L. J. Li, F. Ming, X. K. Song, L. Ye, D. Wang, Quantumness and entropic uncertainty in curved space-time, Eur. Phys. J. C  {\bf82}, 726 (2022).

\bibitem{tQ41}
Aiham M. Rostom, Essential role of destructive interference in the gravitationally induced entanglement, Fortschr. Phys.  {\bf71}, 2200122 (2023).

\bibitem{tQ42}
Aiham M. Rostom, Optimal settings for amplification and estimation of small effects in postselected ensembles, Ann. Phys. (Berlin)  {\bf534}, 2100434 (2022).

\bibitem{tQ43}
X. Liu, W. Liu, Z. Liu, J. Wang, Harvesting correlations from BTZ black hole coupled to a Lorentz-violating vector field,
arXiv: 2503.06404.

\bibitem{tQ44}
W. Liu, C. Wen, J. Wang, Lorentz violation alleviates gravitationally induced entanglement degradation, J. High Energ. Phys. {\bf01}, 184 (2025).



\bibitem{Q41}
D. R. Brill and J. A. Wheeler, Interaction of neutrinos and gravitational fields, Rev. Mod. Phys. {\bf29}, 465 (1957).


\bibitem{Q42}
T. Damoar, R. Ruffini, Black-hole evaporation in the Klein-Sauter-Heisenberg-Euler formalism, Phys. Rev. D {\bf14}, 332 (1976).

\bibitem{Q43}
D. E. Bruschi, J. Louko, E. Mart\'{\i}n-Mart\'{\i}nez, A. Dragan, I. Fuentes, Unruh effect in quantum information beyond the single-mode approximation, Phys. Rev. A {\bf82}, 042332 (2010).

\bibitem{Q44}
Jieci Wang, Jiliang Jing, Heng Fan, Quantum discord and measurement-induced disturbance in the background
of dilaton black holes, Phys. Rev. D {\bf90}, 025032 (2014).

\bibitem{66Q44}
S. M. Wu, H. Y. Wu, Y. X. Wang, J. Wang, Gaussian tripartite steering in Schwarzschild black hole, Phys. Lett. B {\bf865},  139493 (2025).

\end{thebibliography}
\end{document}